\documentclass[pra,twoside,amsfonts,amssymb,amsmath,titlepage,floatfix,twocolumn,showpacs,superscriptaddress]{revtex4} 

\usepackage{amsmath}        
\usepackage{latexsym}
\usepackage{bm}
\usepackage{graphicx}
\usepackage{amsthm}
\newtheorem{thm}{Theorem}

\newtheorem{cor}{Corollary}
\newtheorem{obs}{Observation}
\usepackage{color}

\renewcommand{\phi}{\varphi}

\newcommand{\ket}[1]{|#1 \rangle}
\newcommand{\bra}[1]{\langle #1|}

\newcommand{\be}{\begin{equation}}
\newcommand{\ee}{\end{equation}}
\newcommand{\bea}{\begin{eqnarray}} 
\newcommand{\eea}{\end{eqnarray}}

\newcommand{\Int}{\mathbb{Z}} 
\newcommand{\Nat}{\mathbb{N}} 
 
\newcommand{\Real}{\mathbb{R}} 
\newcommand{\Comp}{\mathbb{C}} 
\newcommand{\Set}{\mathbb{S}} 
\newcommand{\Logic}{\mathbb{L}}

 \newcommand{\tB}{\tilde{B}} 

\begin{document} 
\title{Communication in quantum networks of logical bus topology} 

\author{T. Brougham} 
\affiliation{Department of Physics, FJFI \v CVUT v Praze, B\v rehov\'a 7, 115 19 Praha 1, Star\'e M\v{e}sto, Czech Republic}
\author{G.~M.~Nikolopoulos}
\affiliation{Institute of Electronic Structure and Laser, FORTH, P.O. Box 1527, Heraklion 71110, Crete, Greece}
\author{I. Jex} 
\affiliation{Department of Physics, FJFI \v CVUT v Praze, B\v rehov\'a 7, 115 19 Praha 1, Star\'e M\v{e}sto, Czech Republic}
 
\date{\today}

\begin{abstract} 
Perfect state transfer (PST) is discussed in the context of passive quantum 
networks with logical bus topology,  where many logical nodes communicate using 
the same shared media, without any external control. The conditions under which, 
a number of point-to-point PST links may serve as building blocks for the design 
of such multi-node networks are investigated. 
The implications of our results are discussed in the context of various Hamiltonians 
that act on the entire network, and are capable of providing PST between the logical 
nodes of a prescribed set in a deterministic manner.
\end{abstract} 
 
\pacs{03.67.Hk, 
  03.67.Lx 
} 

\maketitle 
 
\section{Introduction} 
Network topology is the study of the geometric arrangement of the various 
elements in a network (e.g., links, nodes, etc.) \cite{book1}. 
Physical topologies describe how different sites are coupled to each other,  
whereas logical topologies refer to the flow of information between the  
various sites in a particular network. 
The simplest logical topology one may consider is the so-called point-to-point (PP)  
topology where one node (source) is connected directly to another (destination) node.  
The logical network topology is not necessarily the same as the physical layout, since the former is   
bound to the protocols used for communication.
For instance, there might exist intermediate physical connections between the  
source and the destination nodes that do not change the PP 
character of the logical topology, i.e., any message originating from  
the source is intended for the destination node only.  

A direct generalization of the point-to-point logical topology to multiple  
nodes is the so called bus topology, which enables a number of logical nodes to 
communicate using the same shared media. Physically speaking, various  
nodes are connected to a common wire (known as backbone or bus). Whenever 
a node wants to communicate with another node, it sends the message over the 
common channel. In contrast to the PP topology, the message is  
visible and, in principle, accessible by all the logical nodes,  
although the intended recipient is actually the one that accepts and processes  
the message \cite{remark1}. 
 
Recently, many of the questions typically addressed in conventional  
networks have been transferred to a quantum setting, where information is  
imprinted onto the state of quantum systems, and the quantum wires  
are fully compatible with the hardware in a particular quantum computing  
implementation \cite{newref}. 
Typically, quantum networks are discrete, consisting of a
number of coupled quantum objects (physical nodes), a subset of which 
can be the logical nodes. 
Passive quantum networks, where the nodes have fixed energies and 
are permanently coupled, are of particular interest since they do not require any external control 
associated with the application of multiple gates and/or measurements. 
In this spirit, various Hamiltonians have been proposed which  
allow for PST from one node to another 
\cite{sss05,shore,phe04,NPL04,CDEKL04,KS05,YB05,Kay06}. 
In such a PP logical topology, the other nodes provide 
the backbone (or bus), for the communication between the source and the destination 
nodes. The excitation is shared among all of the  
nodes throughout its evolution, and it is only after a well-defined time   
that it can be  localized with high probability at the destination node.
 
This paper goes beyond PP logical topologies, placing the problem of PST in a 
much broader context involving passive networks with logical bus topology.   
Many logical nodes need to communicate with each other using the same shared media,  
and our task is to engineer Hamiltonians, which are capable of transferring 
{\em successively} a quantum state  from the source node, to each one of the other 
logical nodes, in a perfect and deterministic manner  \cite{remark2}. 
This is in contrast to previous studies on multiport  
network topologies \cite{control,WLKGGB07,Y07,GMN08,TG09} where some kind of  
external control is required, to direct selectively the signal to one of 
the destination nodes only. In our case, the passive network 
evolves under the influence of a single time-independent Hamiltonian and, 
in contrast to PP topology, 
the signal is  accessible by all the logical nodes; 
albeit at different times.  

In the following section we formulate the problem under consideration, 
and describe briefly the methodology we adopt throughout this work.

\section{Formalism} 
\label{SecII} 
The system under consideration pertains to a network consisting of $d$ permanently 
coupled physical sites labeled by $\{0, 1, \ldots, d-1\}\equiv \Set_d$, 
and let $\Logic$ denote the set of logical nodes with $\Logic\subseteq\Set_d$. 
The label that we give to each node is arbitrary and does not 
necessarily correspond to any physical ordering of the network.  
For example, the physical nodes 0 and $d-1$ could be directly coupled to 
each other physically.  Similarly, if our network was arranged in a linear chain, then the physical 
nodes 0 and $d-1$ need not be the first and last nodes respectively.  

To recapitulate our formalism, it is sufficient to restrict this section to PP logical networks 
setting $\Logic=\{l_{\rm s},l_{\rm d}\}$, with $l_{\rm s(d)}\in\Set_d$ 
denoting the source(destination) nodes, respectively. 
All the nodes are in the ground state $\ket{\oslash}^{\otimes d}$ and at time $t=0$ 
the source node is prepared in some state $\ket{\psi}$. The problem of PST pertains to the 
quest of Hamiltonians, which 
allow for the transfer of the state to the destination node  $l_{\rm d}$
with unit efficiency after a well-defined time $\tau$. 

Depending on the particular implementation under consideration, the state may 
involve various degrees of freedom besides the position of the excitation 
(e.g., spin, angular momentum, etc). 
Following previous work in the field  
\cite{sss05,shore,phe04,NPL04,CDEKL04,KS05,YB05,Kay06}, we assume 
that all of these additional degrees of freedom 
are preserved throughout the evolution of the system.   
Moreover, we are interested in Hamiltonians which 
preserve the total number of excitations in the system, and we thus focus on 
the transfer of a single excitation from the source to the destination node, 
within the one-excitation Hilbert subspace.  
The corresponding (computational) basis is denoted by 
$\left \{\ket{\kappa}~|~\kappa\in\Set_d\right \}$, where $\ket{\kappa}$ indicates 
the presence of the excitation at the site $\kappa$, i.e., 
$\ket{\kappa}\equiv\ket{\oslash}^{\otimes(\kappa-1)}|\psi\rangle|\oslash\rangle^{\otimes(d-\kappa)}$.
The problem is thus reduced to the quest 
of PST Hamiltonians, which perform the transformation 
$\ket{l_{\rm s}}\to\ket{l_{\rm d}}$ at time $t=\tau$. 

It has been shown \cite{KNJ07} that, for a large class of PST Hamiltonians, the associated unitary 
evolution $\hat{U}$ leads to a permutation matrix at time $\tau$ that permutes 
the source and destinations nodes i.e., 
\bea
\hat{U}(\tau)\equiv e^{-{\rm i}\hat{H}\tau} = \hat{\Pi}, \label{eq:1eq} 
\eea 
where 
\bea 
\hat{\Pi} = \ket{l_{\rm d}}\bra{l_{\rm s}}+\hat{\Pi^\prime}, \label{eq:generalP} 
\eea 
is a permutation in the single excitation sector i.e., the 
submatrix $\Pi^\prime$ is also a  permutation in the computational basis.  
For instance, it is known that a one-dimensional (1D) 
spin chain with a judiciously chosen $XY$ Hamiltonian,   
can be used to perfectly transfer a state from one end of the chain to the 
other \cite{newref,NPL04,CDEKL04}.  The success of this PST protocol is due 
to the fact that the Hamiltonian generates the permutation 
$\hat\Pi=\sum_n{|n\rangle\langle d-n|}$ \cite{KNJ07}.     
For a network with $d$ sites and PP logical topology, there are $(d-1)!$ different 
permutations of the form (\ref{eq:generalP}), and for each one of 
these one can construct the corresponding class of Hamiltonians satisfying
Eq. (\ref{eq:1eq}) (and are thus suitable for PST between the prescribed nodes).

In general, $\hat\Pi$ can be decomposed into disjoint cycles $\hat\Pi_i$ \cite{cycledef}, 
i.e., $\hat \Pi=\sum_i{\hat \Pi_i}$,  and let 
$\Omega_i$ denote the set of sites (and thus basis states) permuted by $\hat\Pi_i$. 
The sets $\{\Omega_i\}$ are disjoint, while the dimensions of the support of each cycle 
$\hat\Pi_i$ is the cardinality of the corresponding set denoted by $d_i$. 
The spectrum of each cycle $\hat\Pi_i$ is nondegenerate and let $|v_{i}^{(\lambda_n)}\rangle$ 
be the eigenvector corresponding to an eigenvalue $\lambda_n$
where $\hat\Pi_i|v_{i}^{(\lambda_n)}\rangle=\lambda_{n}|v_{i}^{(\lambda_n)}\rangle$, with 
\bea
\ket{v_{i}^{(\lambda_n)}}=\frac{1}{\sqrt{d_i}}\sum_{\kappa\in\Omega_i}\lambda_n^{\zeta_\kappa}\ket{\kappa},
\label{eq:McycleEVector} \eea  
and
\bea \lambda_n=\exp\left ({\rm i}2\pi\frac{n}{d_i}\right )\quad\textrm{for} \quad n\in\Int_{d_i}\equiv\{0,1,\ldots, d_i-1\}.
\label{eq:McycleSpectrum}\eea 
The elements of the set $\Omega_i$ are considered to be arranged in ascending order, 
and $\zeta_\kappa\in \Int_{d_i}$ is the position of the element $\kappa\in\Omega_i$. 
Hence, for the permutation $\hat\Pi$ the eigenvalue $\lambda_n$ corresponds to 
$\eta_{\lambda_n}$ distinct eigenvectors $\{|v_{i}^{(\lambda_n)}\rangle\}$, 
with $i$ running only on the various cycles having $\lambda_n$ in common.

A class of PST Hamiltonians that satisfy Eq. (\ref{eq:1eq}) is of the form   
\bea {\cal H}_{\mathbf{x}}=\frac1\tau\sum_{\lambda_n}
\sum_{a=1}^{\eta_{\lambda_n}}\varepsilon_{\lambda_n}^{(a)}
\ket{y_{\lambda_n}^{(a)}}\bra{y_{\lambda_n}^{(a)}},
\label{deg_Ham}
\eea 
where $\varepsilon_{\lambda_n}^{(a)}=-\arg\left (\lambda_n\right )+ 2\pi x_{\lambda_n}^{(a)}$ and 
$\mathbf{x}\in\Int^d\equiv\{(x_{\lambda_0}^{(1)},\ldots,x_{\lambda_0}^{(\eta_{\lambda_0})};x_{\lambda_1}^{(1)},\ldots,x_{\lambda_1}^{(\eta_{\lambda_1})};\ldots)~|~x_{\lambda_n}^{(a)}\in\Int\}$. 
For a given eigenvalue $\lambda_n$, the $\eta_{\lambda_n}$ distinct vectors 
$\{|y_{\lambda_n}^{(a)}\rangle\}$ form an orthonormal basis for the corresponding subspace and are of the form 
\bea
\ket{y_{\lambda_n}^{(a)}}=\sum_{i} \beta_{a,i}^{(\lambda_n)}\ket{v_{i}^{(\lambda_n)}},
\label{y_na}
\eea
with $\beta_{a,i}^{(\lambda_n)}\in\Comp$ and $\sum_i\beta^{(\lambda_n)*}_{a,i}\beta_{a^\prime,i}^{(\lambda_n)}=\delta_{a,a^\prime}$.  
In other words, the rows of the matrix $\tB_{\lambda_n}$ with elements $\beta_{a,i}^{\lambda_n}$ are orthonormal, which implies 
unitarity of $\tB_{\lambda_n}$ and thus orthonormality of its columns as well, i.e., 
\begin{equation} 
\label{orthog} 
\sum_a{\beta_{a,i}^{(\lambda_n)}\beta^{(\lambda_n)*}_{a,i^\prime}}=\delta_{i,i^\prime}. 
\end{equation} 
For a given PST Hamiltonian of the form (\ref{deg_Ham}), i.e., for fixed ${\bf x}$, 
the unitary evolution operator reads
\begin{equation}
\label{app2}
\hat U(t)=\sum_{\lambda_n}\sum_{a=1}^{\eta_{\lambda_n}}\exp\left(-{\rm i}\varepsilon_{\lambda_n}^{(a)} t/\tau \right)
\ket{y_{\lambda_n}^{(a)}}\bra{y_{\lambda_n}^{(a)}},
\end{equation}
with $\ket{y_{\lambda_n}^{(a)}}$ given by Eq. (\ref{y_na}). 

The condition (\ref{eq:1eq}) is rather restrictive since, in general, one may derive 
PST Hamiltonians which do not lead to permutations at time $\tau$, but rather to other 
unitary operations. Despite this fact, it has been demonstrated that the present theoretical 
framework is suitable for engineering of passive networks, and many known PST Hamiltonians 
satisfy condition (\ref{eq:1eq}) \cite{KNJ07}. 
The permutation matrix (\ref{eq:generalP}) guarantees that the net effect of the evolution (\ref{eq:1eq}) 
is the transfer of the excitation from the source to the destination node at time $\tau$. 
The present formalism has the advantage that we do not need to make any assumptions
about the physical topology of the network. Instead infinitely many 
Hamiltonians suitable for PST can be specified within the present theoretical
framework, and this enables us to find the most suitable Hamiltonian for a given 
physical realization and topology.  
Finally, there is no requirement for the initial overall state of the network 
(minus the source node) to have a 
specific form i.e., to be pure or mixed state \cite{initnote}.  
Nevertheless, the source node is considered to be initially decorrelated from the rest 
of the the network, 
which can be prepared in any state \cite{decornote}. Moreover, for the sake of simplicity, 
throughout this work we assume that all the nodes but the source, 
are initially in their ground state.   

We turn now to discuss the problem of PST in the context of logical networks with bus topology. 
In the classical theory of networks \cite{book1} one often assumes that the logical nodes 
have an equivalence property that is, no matter which logical node the signal starts at, it will be 
transferred about the predetermined logical network. Similarly, throughout this work we focus on 
passive quantum networks with equivalent logical nodes, and thus the PST (flow of information) 
is restricted within the predetermined logical network, irrespective of the source of the state.  
In the following section we investigate the conditions under which multiple point-to-point PST links 
can be concatenated to yield PST between multiple nodes.

\section{Concatenation of PP PST links} 
\label{sec2} 
Suppose we are interested in transferring an excitation between three different physical 
sites. For the sake of convenience we shall label these three logical nodes as
$\{l_0, l_1, l_2\}$, regardless of their physical proximity.  
Let us also assume that we have solved the problem of pairwise PST, i.e.,  we have two 
Hamiltonians $\hat h_1$ and $\hat h_2$, which generate the unitary transformations 
$\hat U_1$ and $\hat U_2$, respectively, such that $\hat U_j$ perfectly transfers the state 
from the node $l_0$ to $l_{j}$, i.e., 
\be
\hat U_j|l_0\rangle=e^{{\rm i}\phi_j}|l_{j}\rangle,~\textrm{for}~\phi_j\in\Real.
\label{transf}
\ee
Our task is to investigate the conditions under which 
we can define a single time-independent Hamiltonian ${\hat H}$, which acts on the entire 
network and generates both of the unitaries at different times, 
i.e., $\exp(-{\rm i}\hat{H}\tau_j)=\hat{U}_j$. In this case, the two unitaries 
(or else the corresponding PP PST links) are compatible, 
and can be used as building blocks for the design of passive networks with more involved logical topologies 
pertaining to the prescribed nodes $\{l_j\}$.

The following theorem provides necessary and sufficient conditions for the compatibility 
of two unitary operators.
\begin{thm} 
Two different unitary operators $\hat U_1$ and $\hat U_2$ are compatible if and only if 
\begin{itemize} 
\item[(i)] There exist 
$\tau_1,\tau_2\in(0,\infty)~:~\hat U_1^{\tau_2}=\hat U_2^{\tau_1}~\textrm{for}~\tau_1\neq \tau_2$,
and for all $t\in(0,\infty)$ $\hat U_1^t\neq \hat U_2^t$.
\item[(ii)] The two unitary operators commute, i.e. $[\hat U_1,\hat U_2]=0$. 
\end{itemize} 
\end{thm} 

\begin{proof}[{\bf Proof of Theorem 1.}]
We begin by showing that these two conditions are sufficient for
the two unitary operators to be compatible.  We first assume that we have two
Hamiltonians $\hat h_1$ and $\hat h_2$ that generate two unitary operators
$\hat U_1$ and $\hat U_2$, which satisfy conditions (i) and (ii).  
Condition (ii) ensures that the two unitaries have a common eigenbasis 
(with eigenprojectors $\{\hat E_i\}$) and can be thus simultaneously
diagonalized i.e.,
\begin{eqnarray} 
\hat U_1=\sum_i{\alpha_i\hat E_i},\quad \hat U_2=\sum_i{\gamma_i\hat E_i}. 
\end{eqnarray} 
Condition (i) then implies that $(\alpha_i)^{\tau_2}=(\gamma_i)^{\tau_1}$ for
all $i$,  and setting $\alpha_i=\lambda_i^{\tau_1}$ we also obtain $\lambda_i^{\tau_2}=\gamma_i$.  
We can now define the normal operator $\hat U=\sum_i{\lambda_i\hat E_i}$, where $\hat U^{\tau_1}=\hat
U_1$ and $\hat U^{\tau_2}=\hat U_2$. In order for the operator $\hat U$ to be
unitary we must have that $|\lambda_i|^2=1$ for all $i$. In our case, however, this is 
ensured by the equalities $|\lambda_i|^{2\tau_1}=|a_i|^2=1$ and $|\lambda_i|^{2\tau_2}=|b_i|^2=1$, 
which enable us to choose $|\lambda_i|=1$. Recalling that all
unitary operators can be expressed in the form $\hat U=\exp(-{\rm i}\hat Ht)$, 
where $\hat H$ is Hermitian, it is clear that $\hat H$ will act as a
Hamiltonian on the entire network, generating  the two unitaries 
$\hat U_1$ and $\hat U_2$ at times $\tau_1$ and $\tau_2$, respectively. 

We shall now show that if two different unitary operators 
$\hat U_j$ can be generated from a single  Hamiltonian 
$\hat{H}$, then conditions (i) and (ii) will be satisfied.    
Let $\tau_1$ and $\tau_2$ be two different times and define 
$\hat U_j=\exp(-{\rm i}\hat H\tau_j)$, where $j=1,2$.  By a trivial application of the standard operator 
ordering theorems \cite{br}, we find that 
$\exp(-{\rm i}\hat H\tau_1\tau_2)=\hat U_1^{\tau_2}=\hat U_2^{\tau_1}$. If $\tau_1=\tau_2$,  
then $\hat U_1=\hat U_2$ (which contradicts our initial assumption) 
and thus we must have $\tau_1\ne \tau_2$. It is obvious that no time $t$ can 
exist such that $\hat U_1^t=\hat U_2^t$ for $\hat U_1\ne\hat U_2$.  
Condition (i) is therefore a necessary condition for the two unitary
transformations to be compatible.  Finally, condition (ii) 
is also a necessary condition since $[\exp(-{\rm i}\tau_1\hat H),\exp(-{\rm i}\tau_2\hat H)]=0$.
\end{proof}    
 
The previous discussion pertains to networks with three 
logical nodes $\{l_0,l_1,l_2\}$ and the PST from $l_0$ to $l_{j}$ occurs at 
well-defined time $\tau_j$, with $\tau_2>\tau_1$. 
It is straightforward to extent the notion of compatibility to larger sets 
of unitary operators $\{\hat U_j\}$, and thus to generalize the previous 
discussion to passive networks with $|\Logic|>3$ logical nodes. 

\begin{cor}
The unitary operators of a given set $\{\hat U_j~|~\hat U_j\ne \hat
U_{j^\prime}~\textrm{for}~j\ne j^\prime~\textrm{and}~j,j^\prime\in\Nat\}$ are compatible iff 
any pair of these operators satisfies the conditions of theorem 1.
\end{cor}

If the compatibility conditions are satisfied, there exists a single Hamiltonian to implement 
the unitary operators $\{\hat U_j\}$ at well-defined distinct times 
$\{\tau_j\in(0,\infty)~|~\tau_j>\tau_{j^\prime}~\textrm{for}~j>j^\prime\}$.  

The above theorem and corollary are very general and applicable to various kinds of 
PP PST links, including the schemes of Refs. \cite{shore,sss05,phe04,NPL04,CDEKL04,KS05,YB05,Kay06}. 
To gain, however, further insight into the problem of concatenating a number of given PST links, 
and the possible limitations one may face while looking for solutions, we have to become 
more specific. 
To this end, we assume that one of the unitaries, let us say $\hat{U}_2$, is a permutation matrix 
$\hat \Pi$ that permutes the logical nodes $l_0$ and $l_2$.  
In any case, this choice cannot be considered very restrictive since many known PST schemes can 
be obtained in this context \cite{KNJ07}.
The network is initially in the state $\ket{l_0}$ and in order for $\hat{\Pi}$ and $\hat{U}_1$ to 
be compatible, they must satisfy the conditions of theorem 1. As we will see now, however, the 
structure of the permutation $\hat \Pi$ (i.e., whether or not it can be decomposed into smaller 
closed cycles) implies additional constraints on the logical networks one may consider. 

Suppose that $\hat\Pi$ can be decomposed into closed subcycles $\{\hat\Pi_j\}$ that cannot be 
decomposed further. By definition, $\hat{\Pi}$ permutes the nodes $l_0$ and $l_2$ and thus, these 
sites must belong to the same cycle that we denote by $\hat\Pi_0$. 
The question is whether $l_1$ must also belong to the same cycle, in order 
for the PST within the logical network $\{l_0,l_1,l_2\}$ to be possible.
  
\begin{thm}
Let $\hat H$ be a Hamiltonian governing the evolution of an excitation in a passive 
quantum network, such that 
$\hat\Pi=\exp(-{\rm i}\hat H\tau_2)$, where $\hat{\Pi}$  permutes the nodes 
$l_0$ and $l_2$.  To achieve PST to an intermediate site $l_1$ at an earlier time $\tau_1<\tau_2$, 
$l_1$ must belong to the same closed cycle of $\hat\Pi$ as $l_0$ and $l_2$.
\end{thm} 

Before we prove this statement let us discuss its physical implications.  
If we want to perfectly transfer the state to several different logical nodes at different times, then all 
of these logical nodes must belong to the same cycle of $\hat\Pi$.  
The choice of logical nodes thus places a restriction on our choice of $\hat\Pi$, and vice-versa.  This will in turn impose physical constraints on the type of Hamiltonians that one can derive 
(e.g., see \cite{note1}).   
On the other hand, for the design of a {\em universal bus}, that transfers the excitation 
successively to every node of the network, $\hat\Pi$ must be a one-cycle permutation (see also 
Sec. \ref{bussection}). 
Theorem 2 also implies that we cannot achieve PST 
between two physical nodes that belong to different cycles of $\hat\Pi$.  
This means that the probability of 
finding the state at a node that is not in the same cycle as the source node $l_0$ 
will be strictly less than one. A question therefore is whether the transfer can be achieved 
with high probability.  
The following observation, which is proved in the appendix \ref{app:ob1}, 
partially answers this question.

\begin{obs}
Let two logical nodes $l_0$ and $l_1$ belong to different closed cycles 
$\hat\Pi_0$ and $\hat \Pi_1$ respectively, of a permutation $\hat\Pi$.  It can be shown that 
\begin{equation} 
\label{bound} 
P_{l_1}(t)\le \frac{1}{d_0d_1}\left(\min\{d_0,d_1\}\right)^2, 
\end{equation} 
where $d_j$ is the dimension of the support of $\hat{\Pi}_j$, and 
$P_m(t)=|\langle m|\hat U(t)|l_0\rangle|^2$ is the probability for the excitation to occupy the node 
$m$ at time $t$, where $\hat U(t)=e^{-{\rm i}\hat H t}$ and $\hat U(\tau)=\hat\Pi$, for some $\tau>0$. 

\end{obs}
Inequality (\ref{bound}) provides a useful upper bound (strictly less than 1) 
on the probability $P_{l_1}(t)$ only when $d_0\neq d_1$.
On the other hand, one may expect $P_{l_1}(t)=1$ only for $d_0=d_1$, but our numerical investigations 
show that even in this case the probability is generally much less than 1. 

\begin{proof}[{\bf Proof of Theorem 2.}] Assuming that the nodes $l_0$ and $l_1$ belong to different cycles 
(denoted by $\hat\Pi_0$ and $\hat\Pi_1$, respectively), and given the existence of 
$\hat{\Pi}:\ket{l_0}\to\ket{l_2}$, we will prove by contradiction that there exists no 
unitary operator $\hat U_1$ which is compatible with $\hat \Pi$ and satisfies 
$\hat U_1\ket{l_0}=e^{{\rm i}\phi_1}\ket{l_1}$. It is sufficient, for our purposes, 
to focus on the case of $d_0=d_1$ since, according to inequality (\ref{bound}), 
we cannot achieve PST from $l_0$ to $l_1$ when $d_0\neq d_1$.

Suppose that there exists such a unitary operator $\hat U_1$ which 
satisfies Eq. (\ref{transf}), and can thus be expressed in the form 
\begin{equation} 
\hat U_1=e^{{\rm i}\phi_1}|l_1\rangle\langle l_0|+\hat W, 
\label{1stU1}
\end{equation} 
where unitarity of $\hat U_1$ implies that $\hat W^\dag\ket{l_1}=0$ and $\hat W \ket{l_0}=0$. 
In some sense, this is a first guess of $\hat U_1$, that we can refine further 
using the fact that $\hat\Pi$ and $\hat U_1$ 
are compatible. According to theorem 1, this means that 
$[\hat U_1,\hat\Pi]=0$, which implies that 
$\hat U_1\left(\hat \Pi_0\ket{l_0}\right)=e^{{\rm i}\phi_1}\hat\Pi_1\ket{l_1}$. 
In view of this additional information, Eq. (\ref{1stU1}) can be rewritten as   
$ \hat U_1=e^{{\rm i}\phi_1}\left(\ket{l_1}\bra{l_0}+\hat\Pi_1\ket{l_1}\bra{l_0}\hat\Pi^\dag_0\right)+\hat W_1$, 
where $\hat W_1^\dag\ket{l_1}=\hat W_1^\dag\hat\Pi_1\ket{l_1}=0$ and 
$\hat W_1\ket{l_0}=\hat W_1\hat\Pi_0\ket{l_0}=0$.  
Given, however, that $[\hat U_1,\hat\Pi]=0$, we also have that $U_1$ commutes with any 
power of $\hat{\Pi}$. Hence, in general, for $d_0\geq 1$ we can perform $d_0$ iterations of the 
above refinement with the $j$th iteration pertaining to $[\hat U_1,\hat\Pi^j]=0$. The final result is  
\begin{equation} 
\label{vpro} 
\hat U_1=e^{{\rm i}\phi_1}\sum^{d_0-1}_{j=0}{\hat\Pi^j_1\ket{l_1}\bra{l_0}(\hat\Pi^{\dagger}_0)^j}+\hat W_{d_0-1}, 
\end{equation} 
with $\hat W_{d_0-1}^\dag\ket{\xi_1}=0$ and $\hat W_{d_0-1}|\xi_0\rangle=0$, 
$\forall |\xi_{0(1)}\rangle\in\Omega_{0(1)}$.  
So, in order for the nonsingular $\hat U_1$ to commute with $\hat \Pi$, the map 
$\hat U_1:\Omega_0\mapsto\Omega_1$ must be bijective.  

On the other hand, $[\hat U_1,\hat\Pi]=0$ implies that 
$\hat U_1$ and $\hat \Pi$ are simultaneously diagonalizable, and we can construct a common 
eigenbasis. 
The permutation $\hat \Pi$ has a degenerate spectrum since, by definition, it consists of 
several, let us say $a$, cycles $\{\Pi_i\}$, with $\Pi_0$ and $\Pi_1$ involving the nodes $l_0$ 
and $l_1$, respectively. All the cycles have at least one eigenvalue in common, 
namely $\lambda_0=1$, and let $\ket{v_i^{(\lambda_0)}}$ be the corresponding eigenvector for cycle 
$\hat\Pi_i$, given by Eq. (\ref{eq:McycleEVector}) for $n=0$. In the subspace that is spanned by 
these $a$ distinct eigenvectors, we can construct an orthonormal eigenbasis of 
$\hat \Pi$ (and thus of $\hat U_1$), with elements $\{\ket{y_{\lambda_0}^{(a)}}\}$ given by Eq (\ref{y_na}).

In view of Eqs. (\ref{vpro}), and given that the sets 
$\Omega_j$ are disjoint,  
we have $\hat U_1\ket{y_{\lambda_0}^{(a)}}=\beta_{a,0}^{(\lambda_0)}e^{{\rm i}\phi_1}\ket{v_1^{(\lambda_0)}}
+{\cal O}(\ket{k}~|~k\notin \Omega_0\cup\Omega_1)$.  
Recalling that the state $\ket{y_{\lambda_0}^{(a)}}$ is also an eigenvector of $\hat U_1$ (with eigenvalue 1), 
we find that 
$e^{{\rm i}\phi_1}\beta_{a,0}^{(\lambda_0)}=\beta_{a,1}^{(\lambda_0)}~\forall~a$, and the orthogonality condition (\ref{orthog}) 
yields $\sum_a|\beta_{a,0(1)}^{(\lambda_0)}|^2=0$.  This, however, is not possible because by construction, 
at least one of the vectors $\{\ket{y_{\lambda_0}^{(a)}}\}$ 
will satisfy $|\langle v_{0(1)}^{(\lambda_0)}\ket{y_{\lambda_0}^{(a)}}|\neq 0$ and thus not all 
$|\beta_{a,0(1)}^{(\lambda_0)}|=0$.
We have proved by contradiction, therefore, that PST from node $l_0$ to node $l_1$, 
with the two nodes being parts of different cycles of $\hat\Pi$, is impossible.
\end{proof}

In closing this section, we would like to point out an additional constraint on the structure of the networks 
under consideration, imposed by the assumption of the final unitary being a permutation. 
Using Eqs. (\ref{eq:McycleEVector}),  (\ref{eq:McycleSpectrum}) and (\ref{y_na}), one can express the 
Hamiltonian (\ref{deg_Ham}) in the computational basis, where it is apparent that 
all the physical nodes of the network that belong to the same cycle of the permutation 
must have the same energy. 

\section{Hamiltonians for PST between more than two logical nodes}
\label{severalnodes}
In this section, by employing the theorems of the previous section, we demonstrate how one can construct Hamiltonians that achieve PST within a logical network pertaining to more than two nodes. 

\subsection{The universal bus}
\label{bussection}
Suppose we are looking for Hamiltonians that transfer the excitation successively to each one 
of the $d$ sites of a passive network, within a well-defined time $\tau$.  
We refer to this logical network as the universal bus, since it encompasses the entire physical network.  
The order in which the excitation is transferred 
along the network will be as follows $0\rightarrow 1\rightarrow \ldots\rightarrow d-1$.  
This ordering does not affect the generality of our arguments because 
the labeling of the sites is arbitrary.  We do not allow, however, for the same site to get occupied twice 
within the prescribed time of the transfer. 

According to theorem 2, the Hamiltonian of a universal bus can be associated only with 
a one-cycle permutation \cite{note1}.  
A case in point is the permutation 
\begin{equation}
\label{permd}
\hat \Pi_{\rm ub}=|d-1\rangle\langle 0|+\sum_{m\in\Set_d^*}{|m\rangle\langle m+1|},
\end{equation}
with $m\in\Set_d^*\equiv \Set_d\setminus \{0,d-1\}$, and its eigenvalues and eigenvectors 
given by Eqs. (\ref{eq:McycleSpectrum}) and 
(\ref{eq:McycleEVector}), respectively  for $d_i=d$ and $\Omega_i=\Set_d$. 
The class of Hamiltonians that transfer the excitation from node 0 to 
node $d-1$ at time $\tau$, and the corresponding unitary transformation 
can be obtained from Eqs. (\ref{deg_Ham}) and (\ref{app2}) respectively, after dropping the 
index $a$ and the inner summations.
An important point to note is that any choice for the vector 
${\bf x}=(x_{\lambda_1},\ldots,x_{\lambda_n})$, 
with $n\in\Set_d$, will lead to a Hamiltonian that satisfies $\exp(-{\rm i}\hat H\tau)=\hat\Pi_{\rm ub}$, 
but we are interested in the choices that transfer the state to 
every node of the network.

For the transfer of the excitation from node 0 to node $m\in\Set_d^*$, 
the corresponding matrix element is given by 
\bea
\label{matrixelement}
\langle m|\hat U(t)|0\rangle
&=&\frac{1}{d}\sum_{n=0}^{d-1}{\exp\left [\frac{2\pi {\rm i}n(t'+m)}{d}-2\pi {\rm i} x_{\lambda_n}t'\right ]},
\eea  
where $t'=t/\tau$ and $m\in\Set_d^*$. 
To achieve PST at time $t$, we ask for vectors ${\bf x}$, such that 
$\langle m|\hat U(t)|0\rangle=e^{{\rm i}\phi_m}$, or equivalently 
\begin{equation}
\label{mucondition}
\exp\left [\frac{2\pi {\rm i}n(t'+m)}{d}-2\pi {\rm i}x_{\lambda_n}t'\right ]=e^{{\rm i}\phi_m},~ \forall~n\in\Set_d.
\end{equation}
Instead of trying to find all of the possible solutions to this problem, we will instead 
look for a simple, but infinite class of solutions.  Let us impose the restriction 
that the occupancy of the $m$th node must occur at time $\tau_m=m\tau/(d-1)$.  Then, it is 
straightforward to show that one class of solutions to (\ref{mucondition}) is
\begin{equation}
\label{busspectrum}
x_{\lambda_n}=c_m+n+(d-1)f(n),
\end{equation}
where $c_m\in\Real$ and $f:\Set_d\mapsto\Int$.  One can easily verify that Eq. 
(\ref{busspectrum}) leads to the state being perfectly transferred to the $m$th node at 
time $\tau_m$.  Different choices for $f(n)$ and $c_m$ will lead to different 
spectra for the Hamiltonian, and to different dynamics during the process of 
transferring the state between any two nodes of the network.  
We have thus found an infinite class of universal-bus Hamiltonians, for passive quantum 
networks with resonant physical nodes. This is in contrast to other solutions for ``all-to-all'' 
networks, which require to switch on and off couplings or energy shifts for the nodes, in order to 
achieve PST within a prescribed set of logical nodes \cite{control}.

\subsection{Transfer to a subset of nodes}
\label{examples}
Let us now consider logical networks with bus (but not universal) topology i.e., the set of 
logical nodes is $\{l_j\}=\Logic\subset\Set_d$.
We will construct Hamiltonians, which are capable of transferring successively an excitation 
initially occupying one of the logical nodes, to each one of the other logical nodes in a perfect 
and deterministic manner. As before, one can look for solutions where the occupation times for 
any two logical nodes $l_j$ and $l_{j^\prime}$, with $j>j^\prime$, satisfy 
$\tau_j>\tau_{j^\prime}$. We can impose here an additional constraint namely, PST should occur  
for the nodes of the logical network only. This last requirement automatically excludes the 
universal bus, since it will transfer the excitation successively to any node of the physical 
network. 

Consider a network consisting of five physical nodes (i.e.,  $d=5$),  and three logical nodes 
$\Logic=\{0,2,4\}$. 
According to theorem 2, if we are looking for PST Hamiltonians, which at time $\tau$ lead to a 
permutation $\Pi$ that permutes the nodes 0 and 4, then all three logical nodes have to 
belong to the same cycle of $\Pi$.  Moreover, to restrict the flow of information within the logical network only, 
we should avoid choosing permutations for which nonlogical nodes are in the same cycle as the logical ones \cite{note2}. 

A permutation that satisfies the above requirements is 
\begin{equation}
\label{example1}
\hat\Pi=|4\rangle\langle0|+|0\rangle\langle2|+|2\rangle\langle4|+|1\rangle\langle3|+|3\rangle\langle1|,
\end{equation}
which can be decomposed 
into two cycles $\hat\Pi_0= |4\rangle\langle0|+|0\rangle\langle2|+|2\rangle\langle4|$ 
and $\hat\Pi_1=|1\rangle\langle3|+|3\rangle\langle1|$.  The eigenvectors and eigenvalues of these 
cycles can be obtained from Eqs. (\ref{eq:McycleEVector}) and (\ref{eq:McycleSpectrum}) respectively,
for $\Omega_0=\{0,2,4\}$,  $\Omega_1=\{1,3\}$, $d_0=3$ and $d_1=2$.
The spectrum of $\hat\Pi$ is degenerate, since the two cycles have a common eigenvalue 
$\lambda_0$, while an eigenbasis can be constructed along the lines of Sec. \ref{SecII}.
Taking into account the  orthogonality condition one obtains 
\begin{subequations}
\label{ebasis1}
\begin{eqnarray}
&&|y_{0}^{(1)}\rangle=\alpha|v_0^{(0)}\rangle+\beta|v_1^{(0)}\rangle,\\
&&|y_0^{(2)}\rangle=\beta^*|v_0^{(0)} \rangle-\alpha^*|v_1^{(0)}\rangle,\\
&&|y_{\pi}^{(1)}\rangle=|v_1^{(\pi)}\rangle,\\
&&|y_{{k2\pi/3}}^{(1)}\rangle=|v_0^{({k2\pi/3})}\rangle~{\rm for}~k=1,2,
\end{eqnarray}
\end{subequations}
where $\beta_{1,0}^{(1)}=\alpha$, $\beta_{1,1}^{(1)}=\beta$,  $\beta_{2,0}^{(1)}=\beta^*$, 
$\beta_{2,1}^{(1)}=-\alpha^*$, and $|\alpha|^2+|\beta|^2=1$.  Note that, to suppress notation,  
the vectors as well as the components of the integer vector ${\bf x}$ in the following, 
are labeled according to $\arg(\lambda_n)$, and not the actual eigenvalue $\lambda_n$.

A class of PST Hamiltonians that transfer the excitation from node 0 to node 4 
at time $\tau$ is given by Eq. (\ref{deg_Ham}), and the corresponding evolution operator 
is of the form (\ref{app2}).  
We shall now investigate the choices of the spectrum $\{\varepsilon_{\lambda_n}\}$, and thus 
of the integer vector ${\bf x}$, which enable us  
to achieve also PST from node 0 to node 2, at an earlier time $t<\tau$.  
To this end, we focus on the matrix element 
$\langle 2|\hat U(t)|0\rangle$ for which, using Eq. (\ref{ebasis1}), we find 
\begin{eqnarray}
\label{ematrix1}
\langle 2|\hat U(t)|0\rangle&=&\frac{|\alpha|^2}{3}\exp(-2\pi {\rm i}x_{0}^{(1)}t')\nonumber\\
&&+\frac{1}{3} \exp\left [\frac{2\pi {\rm i}(t'+1)}{3}-2\pi {\rm i}x_{{\rm 2\pi/3}}^{(1)}t'\right ]
\nonumber\\
&+&\frac{|\beta|^2}{3}\exp(-2\pi {\rm i}x_{0}^{(2)}t')\nonumber\\
&&+\frac{1}{3} \exp\left [\frac{4\pi {\rm i}(t'+1)}{3}-2\pi {\rm i} x_{{\rm 4\pi/3}}^{(1)}t'\right ],
\end{eqnarray}
where $t^\prime=t/\tau$.
We also have for the matrix elements pertaining to non-logical nodes
\begin{eqnarray}
\label{notransfer1}
\langle 1|\hat U(t)|0\rangle&=&\langle 3|\hat U(t')|0\rangle=
\frac{\beta\alpha^*}{\sqrt{6}}\exp(-{\rm i} 2\pi x_{0}^{(1)} t')\nonumber\\
&&-\frac{\alpha^*\beta}{\sqrt{6}}\exp(-{\rm i} 2\pi x_{0}^{(2)} t').
\end{eqnarray}
Equation (\ref{ematrix1}) shows that we are free in choosing $x_{\pi}^{(1)}$, as this will not 
affect the probability of transferring the excitation to node 2. 

To proceed further we can specify a time at which the state will be transferred from node 0 to 
node 2, let us say $\tau_2=\tau/2$ or equivalently $t^\prime=1/2$.  
Hence, we ask for $\langle 2|\hat U(\tau_2)|0\rangle=e^{{\rm i}\phi_2}$.  
This requirement will constrain the allowed values of the integers $\{x_{\lambda_n}^{(a)}\}$, 
that appear in Eq. (\ref{ematrix1}).  We can distinguish between two cases. 
(i) For $|\alpha|=0$ or 1, either $|\alpha|^2\exp(-2\pi {\rm i} x_{0}^{(1)}t')/3$ 
or $|\beta|^2\exp(-2\pi {\rm i} x_{0}^{(2)}t')/3$ will be zero and thus, without loss of generality, 
we can set $x_{0}^{(1)}=x_{0}^{(2)}$. 
Working similarly to Sec. \ref{bussection}, we find that the transfer 
is facilitated by choosing 
\be
x_{\nu 2\pi/3}^{(1)}=c+\nu+2f(\nu), \textrm{for}\, \nu\in\{0, 1, 2\}\equiv\Int_3,
\label{cond2}
\ee 
where $c$ is a constant and $f:\Int_3\mapsto\Int$. 
(ii) For $0<|\alpha|<1$,  in addition to $|\langle 2|\hat U(\tau_2)|0\rangle|=1$ we require that 
$|\langle 1(3)|\hat U(\tau_2)|0\rangle|=0$, which implies that $x_0^{(1)}=x_0^{(2)}+q$ for even
integers $q\in\Nat$.  As in case (i), it can be readily checked that a suitable choice 
for the other components of ${\bf x}$ is provided by Eq. (\ref{cond2}). It should be emphasized 
also that according to observation 1, the occupation probabilities of the non-logical nodes 
(1 and 3), cannot exceed 2/3 for any $t<\tau$, and thus PST occurs within the logical network only. 
In the appendix \ref{app:spin}, we discuss a possible physical ralization of this PST
Hamiltnian in the context of spin chains. 
Finally, it is straightforward to generalize the previous derivation to a case where the PST from 
node 0 to node 2 has to occur at a time $\tau_m=\tau/m$, for $m>2$.

\begin{figure*}
\includegraphics[width=15.cm]{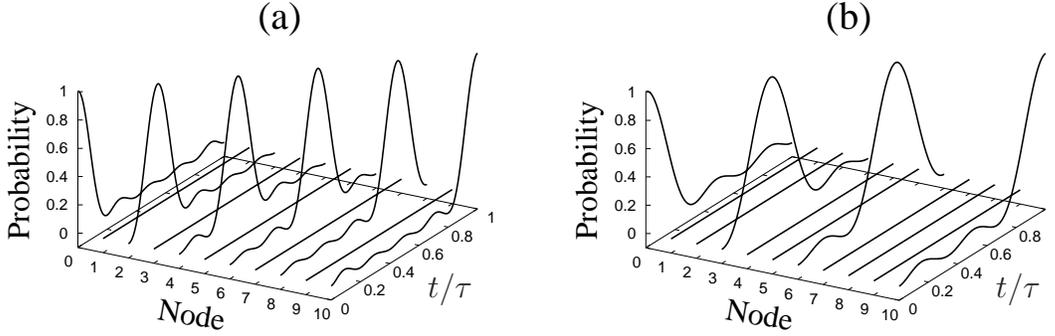}
\caption{PST in a network of logical bus topology, with logical nodes: 
$\{0, 2, 4, 6, 8, 10\}$ (a), and $\{0, 3, 6, 10\}$ (b). 
Each plot pertains to a network consisting of eleven physical sites, and the 
occupation probability distribution is plotted as a function of time.}
\label{fig1}
\end{figure*}

For networks with $d>5$ nodes, it is rather difficult to obtain analytic results, 
but one can resort to numerical investigations. In Fig. \ref{fig1}  
we present results pertaining to a network consisting of 11 physical nodes, and two different 
choices of logical nodes.  
For Fig. \ref{fig1}(a), the set of logical nodes is $\Logic_{\rm a}=\{0, 2, 4, 6, 8, 10\}$, 
whereas for Fig. \ref{fig1}(b), $\Logic_{\rm b}=\{0, 3, 6, 10\}$.   
To construct Hamiltonians that allow for PST within these sets of logical nodes, one has to 
work as before.
Following theorem 2 and observation 1, in both cases we have chosen  
a permutation which has a cycle that contains only the logical nodes e.g., 
$\Pi_{\rm a}=\ket{10}\bra{0}+\ket{8}\bra{10}+\ket{6}\bra{8}+\ket{4}\bra{6}+\ket{2}\bra{4}+\ket{0}\bra{2}+\ket{9}\bra{1}+\ket{7}\bra{9}+\ket{5}\bra{7}+\ket{3}\bra{5}+\ket{1}\bra{3}$,
and $\Pi_{\rm b}=\ket{10}\bra{0}+\ket{6}\bra{10}+\ket{3}\bra{6}+\ket{0}\bra{3}+\ket{9}\bra{1}+\ket{8}\bra{9}+\ket{7}\bra{8}+\ket{5}\bra{7}+\ket{4}\bra{5}+\ket{2}\bra{4}+\ket{1}\bra{2}$. 
Each one of these permutations consists of two cycles and let $d_{0(1)}$ denote the cardinality of 
the cycle containing the node $0(1)$, respectively.
The class of Hamiltonians that transfer the excitation from node 0 to node 10 
at time $t=\tau$  is given by Eq. (\ref{deg_Ham}), and one can choose the integer vector ${\bf x}$, 
and thus the spectrum $\{\varepsilon_{\lambda_n}\}$, so that the excitation is transferred successively
to the other nodes of the logical network $\Logic_{\rm a(b)}$ 
at well defined times $\tau_{\rm a(b)}^{(m)}<\tau$, for $m\in\Logic_{\rm a(b)}\setminus\{0,10\}$. 
When asking for these times to be equally spaced i.e., 
$\tau_{\rm a(b)}^{(m)}=m\tau/(|\Logic_{\rm a(b)}|-1)$, one can choose: 
(a) $x_{2\pi n_{0}/d_0}^{(1)}=11-n_0$, $x_{2\pi n_{1}/d_1}^{(1)}=5-n_1$, $x_{0}^{(1)}=11$, and 
$x_{0}^{(2)}=5$;
(b) $x_{2\pi n_{0}/d_0}^{(1)}=11-n_0$, $x_{2\pi n_{1}/d_1}^{(1)}=11-n_1$, and $x_{0}^{(1,2)}=11$, 
where $n_{\rm 0(1)}\in\Int_{d_0(d_1)}\setminus\{0\}$. 
The evolution of the corresponding occupation probability distributions (defined in observation 1) 
are depicted in Figs. \ref{fig1}(a-b).

\section{Summary}

We have extended the problem of PST to passive quantum networks of bus logical topology, where 
the information flow is restricted within a prescribed set of logical nodes. 
The excitation evolves under the influence of a single Hamiltonian which acts on the entire network, 
and is transferred successively to each one of the logical nodes in a perfect and deterministic 
manner. 
We have provided necessary and sufficient conditions for the engineering of such networks, 
by concatenating a number of point-to-point PST links of any type. 
Our theory is not subject to any {\em a priori} restrictions on the physical topology of the network, 
or its initial state, and provides new ways for quantum network engineering beyond point-to-point 
logical topology.  

The network engineering has been demonstrated in the context of a set of point-to-point PST links, 
where the transformation associated with one of them is a permutation.  Such an assumption 
automatically imposes certain restrictions on the physical topology of the network, but 
is not very restrictive since many known PST Hamiltonians can be obtained in this framework. 
Although our formalism allows for the construction of Hamiltonians of any kind, which can be 
rather tedious, we restricted our network engineering to Hamiltonians that involve complex 
couplings.
In spin networks, the adjustment of geometric phases is possible by looping around magnetic 
fields along the relevant sections \cite{newref,KayEr05}, while for optical networks one may use 
phase shifters.

A number of interesting questions, such as the design of passive quantum networks with 
other logical topologies, the problem of time-limited PST, and the transfer of 
entangled states (encoded on two or more nodes) in networks of bus topology, deserve 
further investigation.  

\section*{Acknowledgements}
TB and IJ acknowledge support from MSM 684077039 and LC06002 of the Czech Republic. 
GMN acknowledges partial support from the EC RTN EMALI (contract No. MRTN-CT-2006-035369).

\appendix
\section{Proof of observation 1}
\label{app:ob1}
To prove inequality (\ref{bound}), we start from Eq. (\ref{app2}) which provides the operator
$\hat U(t)=\exp(-{\rm i}\hat Ht)$ with $\hat U(\tau)=\hat \Pi$. 
Using the triangle inequality we have for the matrix element of interest 
\begin{equation}
\label{app3}
|\langle l_1|\hat U(t)|l_0\rangle|\le\sum_{\lambda_n,a}{\left|\langle l_1 \ket{y_{\lambda_n}^{(a)}}\bra{y_{\lambda_n}^{(a)}}  l_0\rangle\right|}.
\end{equation}  
Given that $l_{0(1)}\in\hat\Pi_{0(1)}$,  it is clear that the matrix elements 
$|\langle l_{0(1)}\ket{y_{\lambda_n}^{(a)}}|$ vanish unless $\lambda_n$ 
is an eigenvalue of $\hat\Pi_{0(1)}$, respectively.  Let $\sigma$ denote the set of eigenvalues that 
$\hat\Pi_0$ and $\hat\Pi_1$ have in common.  Then, using Eqs. (\ref{eq:McycleEVector}) and 
(\ref{y_na}), we have
\bea
\label{app4}
|\bra{l_1}\hat U(t)\ket{l_0}|&\leq&
\frac{1}{\sqrt{d_1d_2}}
\sum_{\lambda_n\in\sigma}{\sum_a{|\beta_{a,0}^{(\lambda_n)}||\beta_{a,1}^{(\lambda_n)}|}}\nonumber\\
&\leq &
\frac{1}{\sqrt{d_1d_2}}\sum_{\lambda_n\in\sigma}\sqrt{
\sum_a{|\beta_{a,0}^{(\lambda_n)}|^2}\sum_a{|\beta_{a, 1}^{(\lambda_n)}|^2}},\nonumber\\ 
\eea
where the second part is due to Cauchy-Schwarz inequality.
Finally, using Eq. (\ref{orthog}) we obtain 
$|\langle l_1|\hat U(t)|l_0\rangle|\leq \left (\sum_{\lambda\in\sigma}{1} \right )/\sqrt{d_0d_1}$, 
where the summation equals the total number of eigenvalues that $\hat\Pi_0$ and $\hat\Pi_1$ 
have in common, and cannot exceed the total number of eigenvalues of either of the two cycles.  
Hence, we have that 
$|\langle l_1|\hat U(t)|l_0\rangle|\leq \min\{d_0,d_1\}/{\sqrt{d_0d_1}}$,
which when squared leads to inequality (\ref{bound}). 

\section{PST in a spin chain with three logical nodes}
\label{app:spin}
The formalism we have adopted throughout this work is rather general, and can be 
easily adjusted to a particular physical implementaiton of the network. 
For instance, to make the example of Sec. \ref{examples} more concrete, 
let us consider the situation where the network is a spin chain, with $d=5$ and 
$\Logic=\{0,2,4\}$.  

In this case each node possesses two degrees of freedom, corresponding to 
the individual spins being either up or down along a particular 
direction, i.e. $|\pm\rangle$, with $\ket{-}^{\otimes 5}$ denoting the ground state 
of the chain.  The source node $(0)$ is considered to be initially de-correlated from the 
rest of the spins \cite{decornote}, and let the state of the entire chain be  
$\ket{\Psi(0)}=|\psi\rangle_0|-\rangle^{\otimes 4}$, 
where $|\psi\rangle=b_1|+\rangle+b_2|-\rangle$ 
(see also related discussion towards the end of Sec. \ref{SecII}). 
The state of the source site has to be transferred successively to the sites 2 and 4, 
at the prescribed times $\tau_2=\tau/2$ and $\tau_4=\tau$, respectively. 
This means that the state of the entire chain  has to transform according to 
 $\ket{\Psi(0)}\to\ket{\Psi(\tau_2)}\to\ket{\Psi(\tau_4)}$, where 
$\ket{\Psi(\tau_j)}=|\psi\rangle_j|-\rangle^{\otimes 4}$.
The initial state, however, is basically a superposition of the ground 
state and the first-excited state 
of the chain, i.e., $\ket{\Psi(0)}=b_1|+\rangle_0|-\rangle^{\otimes 4}+b_2|-\rangle^{\otimes 5}$.  
Thus, given that the ground state does not evolve in time, for the transfer of 
$\ket{\Psi}$ in time, it is sufficient to consider PST in the one-excitation 
subspace i.e., $\ket{0}\to\ket{2}\to\ket{4}$, where 
\be
\label{eq:trans}
|m\rangle=|-\rangle^{\otimes m}|+\rangle_m|-\rangle^{4-m}.
\ee 

The solution of this problem has been dicussed in Sec. \ref{examples}, in the context of 
a PST Hamiltonian that satisfies Eq. (\ref{eq:1eq})  with $\hat{\Pi}$ 
given by Eq. (\ref{example1}). 
A class of PST Hamiltonians that transfer the excitation from spin 0 to spin 4 
at time $\tau$ is given by Eq. (\ref{deg_Ham}), with $\ket{y_{\lambda_n}^{(a)}}$ 
given by Eq. (\ref{ebasis1}). This class of Hamiltonians can be expressed in terms of 
the the Pauli spin operators $\hat X_j$ and $\hat Y_j$  acting on the $j$th spin, 
with $\hat X_j=\ket{-}_j\bra{+}+ \ket{+}_j\bra{-}$ and 
$\hat Y_j={\rm i}(\ket{-}_j\bra{+}-\ket{+}_j\bra{-})$, using the convention (\ref{eq:trans}).
It is convenient to split the analysis into two different cases i.e., to 
couplings between spins that belong to the same cycle, and couplings between spins that 
belong to different cycles. 

A simple calculation shows that the spins 1 and 3 (non-logical nodes), 
which belong to the same cycle $\hat\Pi_1$, are coupled by a term in the 
Hamiltonian that has the following form
\begin{eqnarray}
\label{coup1}
\frac{1}{4\tau}\left(|\beta|^2\varepsilon^{(1)}_0+|\alpha|^2\varepsilon^{(2)}_0+\varepsilon^1_{\pi}\right)\left(\hat X_1\hat X_3+\hat Y_1\hat Y_3\right).
\end{eqnarray} 
The couplings between the other three spins (logical nodes) that belong to the cycle 
$\hat\Pi_0$, have the form
\begin{equation}
\label{coup2}
\frac{J_{mn}}{2}\left(\hat X_m\hat X_n+\hat Y_m\hat Y_n\right)+\frac{J_{mn}'}{2}\left(\hat X_m\hat Y_n-\hat Y_m\hat X_n\right),
\end{equation}
with $m,n\in\Omega_0$ and $m>n$.  
The expressions for the coupling constants $J_{mn}$ and $J_{mn}'$ are
\begin{eqnarray}
\label{js1}
J_{20}=J_{40}=J_{42}&=&\frac{1}{3\tau}\bigg [|\alpha|^2\varepsilon^{(1)}_0+|\beta|^2\varepsilon^{(2)}_0
\nonumber\\
&&-\frac{\varepsilon^{(1)}_{2\pi/3}+}{2}-\frac{\varepsilon^{(1)}_{4\pi/3}}{2}\bigg ],\\
J_{20}'=-J_{40}'=J_{42}'&=&\frac{1}{2\sqrt{3}\tau}\left[\varepsilon^{(1)}_{2\pi/3}-\varepsilon^{(1)}_{4\pi/3}\right].
\end{eqnarray}

The couplings between spins that belong to different cycles have the same 
form as Eq. (\ref{coup2}), with $m\in\Omega_0$ and $n\in\Omega_1$. The 
corresponding coupling constants are given by 
\begin{equation}
\label{js2}
J_{mn}=\frac{\Re(\alpha\beta^*)}{\sqrt{6}\tau}\left[\varepsilon^{(1)}_0-\varepsilon^{(2)}_0\right],\;J_{mn}'=\frac{\Im(\alpha\beta^*)}{\sqrt{6}\tau}\left[\varepsilon^{(1)}_0-\varepsilon^{(2)}_0\right],
\end{equation}
where $\Re(\alpha\beta^*)$ and $\Im(\alpha\beta^*)$ denote the real and 
imaginary parts of $\alpha\beta^*$ respectively, while $J_{mn}'=-J_{nm}'$ for $m<n$.  
The values of the coupling constants, given by Eqs. (\ref{coup1}), (\ref{js1}) and (\ref{js2}) 
are real and depend on the choice of the integers $x^{(a)}_j$, as well as on the parameters 
$\alpha$ and $\beta$.  The coupling constants 
can thus be adjusted while still obtaining Hamiltonians that perform the desired PST. 
For instance, taking $x^{(1)}_0=x^{(2)}_0$ will result in vanishing couplings 
between spins that belong to different cycles.  
 
Finally, setting for convenience the ground state energy equal to zero, one readily 
obtains for the diagonal terms of the Hamiltonian 
\bea
\langle 0|\hat H|0\rangle&=&\langle 2|\hat H|2\rangle=\langle 4|\hat H|4\rangle\nonumber\\
&&=\frac{1}{3\tau}\left[\varepsilon^{(1)}_0|\alpha|^2+\varepsilon^{(2)}_0|\beta|^2
+\varepsilon^{(1)}_{2\pi/3}+\varepsilon^{(1)}_{4\pi/3}\right]\nonumber \\
\langle 1|\hat H|1\rangle&=&\langle 3|\hat H|3\rangle=\frac{1}{2\tau}\left[\varepsilon^{(1)}_0|\beta|^2+\varepsilon^{(2)}_0|\alpha|^2+\varepsilon^{(1)}_{\pi}\right].\nonumber\\
\eea
It can be seen that the nodes belonging to the same cycle have the same energy, when they are in the state $|+\rangle$.  The energies of each node can be adjusted by changing the values of the integers $x^{(a)}_j$.


\begin{thebibliography}{99}
\bibitem{remark1} In a classical setting, this is achieved by means of the  
so-called internet-protocol (IP) addresses.  
As the message arrives at each node, the corresponding device checks the  
destination address contained in the message to see if it matches its own  
address. If the two addresses match the device processes the message, 
otherwise it does nothing.  Clearly, no addressing is required in a 
point-to-point topology since any data transmitted from one port is intended 
for the other port.
\bibitem{remark2} This is not the same as the quantum cloning and entanglement  
distribution discussed in Refs. \cite{HB02-04,CFMMP04-05},  
where the excitation is distributed among the output ports.
\bibitem{cycledef} A closed cycle is a permutation or sub-permutation, 
which cannot be decomposed further.  For example, the permutation 
$\hat{\Pi}= \ket{3}\bra{0} + \ket{2}\bra{1} + \ket{1}\bra{2} +\ket{0}\bra{3}$ 
is not a closed cycle as we can decompose it 
into two smaller permutations $\hat{\Pi}_0=\ket{0}\bra{3}+ \ket{3}\bra{0}$ and 
$\hat{\Pi}_1=\ket{1}\bra{2}+ \ket{2}\bra{1}$.  
An example of a closed cyle, with more than two elements, is $\Pi= \ket{1}\bra{0} + \ket{2}\bra{1} + \ket{0}\bra{2}$; 
it is clear that this permutation cannot be decomposed further.
\bibitem{initnote} The fact that the current formalism is not contingent on the initial state of 
the network greatly reduces the difficulty in realising PST \cite{FPK08}.
\bibitem{decornote}  The assumption that the source node is initially decorrelated from the rest of 
the network is necessary for an unambiguous definition of PST.
\bibitem{note1} As shown in \cite{KNJ07}, for one-dimensional networks, where each node 
has at most two nearest neighbours, nearest-neighbor interaction Hamiltonians cannot be 
obtained in the context of one-cycle permutations. Hence, in this case 
the Hamiltonian of a universal bus cannot be of a nearest-neighbor type.
\bibitem{note2}
A case in point is the permutation    
$\hat \Pi=|4\rangle\langle 0|+|0\rangle\langle 1|+|1\rangle\langle 2|+|2\rangle\langle 4|+|3\rangle\langle 3|$.
\end{thebibliography}
\end{document}